\documentclass[a4paper,12pt]{article}
\usepackage{graphicx,epsf,a4wide}
\newcommand{\eq}[1]{Eq.~(\ref{#1})}
\author{G. Bonnet \\
        \smallskip
        CEA/Saclay, Service de Physique Th\'eorique\\
        F-91191 Gif-sur-Yvette Cedex, France
}
\title{Solution of Potts-3 and Potts-$\infty$ matrix models with the equations of motion method }
\begin{document}
\begin{titlepage}
\raisebox{-1.2cm}{\includegraphics[width=5.cm]{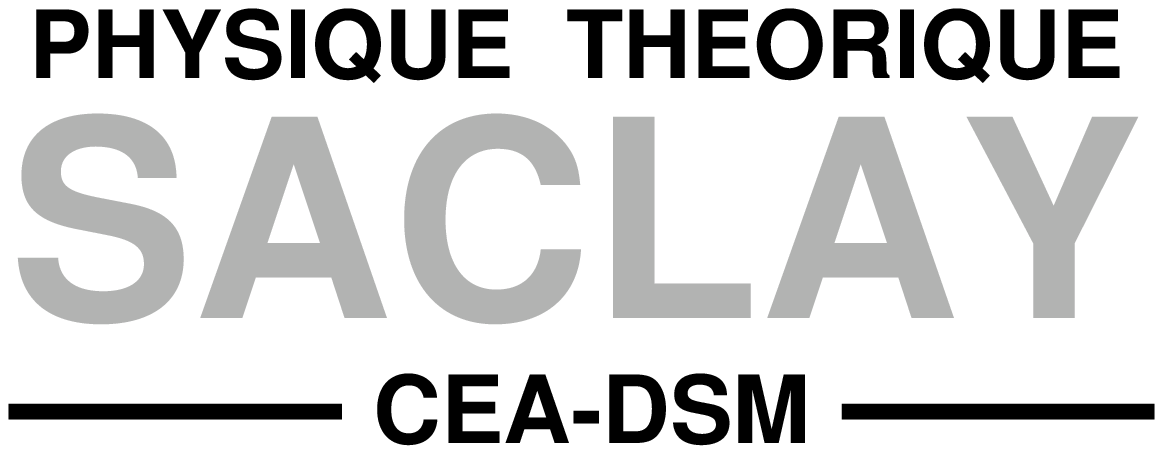}}
\hfill
\begin{minipage}[t]{3.cm}
{\mbox{hep-th/9904058}\\ Saclay T99/020\\}
\end{minipage}
\\[1.cm]
\begin{center}
\textbf{\LARGE Solution of Potts-3 and Potts-$\infty$ 
\\
 Matrix Models  with 
\\\medskip the Equations of Motion Method}
\\[4.ex]
{\large Gabrielle \textbf{Bonnet}\,\footnote{AMN}
\footnote{gabonnet@spht.saclay.cea.fr} }
\\
        \bigskip
        CEA/Saclay, Service de Physique Th\'eorique\\
        F-91191 Gif-sur-Yvette Cedex, France
\end{center}
\bigskip

\begin{abstract}
In this letter, we show how one can solve easily the Potts-3
+ branching interactions
and Potts-$\infty$ matrix models,
by the means of the equations of motion (loop equations).
We give an algebraic equation for the resolvents of these
models, and their scaling behaviour. This shows that the equations of
motion can be a useful tool for solving such models.
\end{abstract}

\end{titlepage}%
\section{Introduction :}
\label{intro}

Random matrices are useful for a wide range of physical problems.
In particular, they can be related to two-dimensional quantum gravity
 coupled to matter fields with a non-zero central charge $C$ \cite{refgen}.
While $C \leq 1$ models are relatively well understood, the $C >1$ domain
remains almost totally unknown : there is a $C=1$ ``barrier''. 
When studying $C \neq 0$ models, we are led to consider multi-matrix models 
\cite{staudacher,eynard}
 which are often non-trivial. One class of difficult matrix models
 corresponds to the $q$-state Potts model (in short: Potts-$q$) on
 a random surface.
This model is a $q$-matrix model where all the matrices are coupled to
each other, thus making difficult the use of usual techniques such as
the saddle point or the orthogonal polynomials method.  
Moreover, the $q \rightarrow 4$ limit corresponds to $C \rightarrow 1$,
thus, by solving Potts-$q$ models, we shall gain a new understanding of
the $C=1$ barrier.

In this letter, we show that, contrary to what was previously thought,
one can use the loop equations to solve the Potts-3 random matrix
model, and we find that the resolvent (which generates many of the 
operators of the problem) obeys an algebraic equation that we write 
explicitely.

We also show that this method applies when one adds branching interactions
(gluing of surfaces, also called ``branched polymers'') \cite{bp}
 and we derive the critical line of this extended model.
The extension to the model with branching interactions and the study of its
phase diagram is necessary to verify \cite{FD}'s conjecture about
 the $C=1$ transition.

Finally, we apply the method to the Potts-$\infty$ matrix model, which
corresponds to $C=\infty$.

As this work was approaching its completion, a paper appeared on the dilute
Potts model \cite{PZJ}, which partially overlaps our present work.
In this article, the author also has an algebraic equation for the
conventional Potts-3 model.
Here, we go further as we consider the Potts-3 + branching interactions model.
Moreover, his method is quite different : while he uses analytical
considerations on the resolvents and large-$N$ techniques, we solve
our model by the loop equations method, which can be extended to
finite $N$ problems and is also more adapted to the use of renormalization
group techniques \cite{higuchi,GBFD}.

\section{The Potts-3 + branching interactions model :}

Let us define :

\begin{equation}
Z=\int \,  d\Phi \, e^{-N^2 V(\Phi)}
\end{equation}

\begin{equation}
V(\Phi)=g  {\makebox{tr} \Phi^3 \over 3 N}+ \psi({ \makebox{tr} \Phi^2 \over 2 N}  , { \makebox{tr} \Phi \delta \Phi \delta \over 2 N} )
\end{equation}

\begin{equation}
\Phi= \left( \begin{array}{ccc}
\Phi_1 & 0 & 0\\
0 & \Phi_2 & 0\\
0 & 0 & \Phi_3\\ 
\end{array} \right) , \,  \delta_{1}= \left( \begin{array}{ccc}
0 & 1 & 0\\
0 & 0 & 1\\
1 & 0 & 0\\
\end{array} \right) , \, \delta_{-1}= \left( \begin{array}{ccc}
0 & 0 & 1\\
1 & 0 & 0\\
0 & 1 & 0\\
\end{array} \right)
\end{equation}

$\delta= \delta_{1} + \delta_{-1}$. We shall
also use later the notation $\delta_{0} = Id .$\\\

\noindent $\Phi$, $\delta_0$, $\delta_1$ and $\delta_{-1}$
 are $(3 N) \times (3 N)$ and $\Phi_1$,
$\Phi_2$, $\Phi_3$  $N \times N$ hermitean matrices.
$\psi$ is a general two-variable function, and will mainly appear
through its partial derivatives $U$ and $c$  with respect to 
${\makebox{tr} \Phi^2 \over 2 N}$ and ${\makebox{tr} \Phi \delta \Phi \delta
\over 2 N}$  respectively. If these are constants, then we
recover the conventional Potts-3 model (no branching interactions).
This model was given partial solution by J.M. Daul 
\cite{daul}, by considering the analytic structure of the resolvents.
He had its critical point  and its
 associated critical exponent. He did not know, however, if the resolvent
 obeyed an algebraic equation. We shall give here the expression of this
equation for the conventional and extended Potts-$3$ model. We also derive the 
critical line of the extended model and check it corresponds to Daul's
result in the particular case of the conventional model.
 \\

Let us note, for convenience :
\begin{equation} 
t_{i_1 i_2 \cdots i_n \Phi^k} ={1 \over 3 N} \langle \makebox{tr}
\delta_{i_1} \Phi \delta_{i_2} \Phi \ldots \delta_{i_n} \Phi^k \rangle
\end{equation}
where $i_1$, $\ldots$, $i_n$ can be $+1$, $-1$ or $0$.
This trace is non-zero if and only if $i_1+\ldots+i_n \equiv 0 \pmod{3}$.
 $\langle \ldots \rangle$ is the expectation value of $(\ldots)$ :
\begin{equation}
 \langle \ldots \rangle= {1 \over Z} \int d\Phi \, (\ldots)\,  e^{-N^2 V(\Phi)} 
\end{equation}
A trace will be said to be ``of degree $m$'' if there are $m$ matrices $\Phi$
in it. For example, the above trace is of degree $k+n-1$.\\\

Let us now use the method of the equations of motion (or loop equations). 
If we make the infinitesimal change of variables in $Z$ :
\begin{equation}
\Phi \rightarrow \Phi + \epsilon \,\delta_{i_1} \Phi\, \delta_{i_2} \ldots 
\Phi \delta_{i_n}
\end{equation}
with $$i_1+i_2+\ldots +i_n \equiv 0  \pmod{3} $$ then we obtain the 
expression of the general 
equations of motion : 
\begin{equation}
\label{eq7}
g \, t_{i_1 \ldots i_n \Phi^2}+U \, t_{i_1 \ldots i_n \Phi} +c\, 
(t_{(i_1+1) i_2 \ldots (i_n-1) \Phi} + t_{(i_1-1) i_2 \ldots (i_n+1) \Phi})
- \sum_{j=1}^{n-1} t_{i_1 \ldots i_j} \, t_{i_{j+1} \ldots i_n} =0
\end{equation}

The first three terms
 come from the transformation of $V(\Phi)$, and the last
one, from the jacobian of the transformation.

\eq{eq7} relates any expectation value of trace containing a quadratic term
(i.e. a $\Phi^2$ term) to expectation values of traces of lower degrees.
The problem is that we do not have any recursion relation for more 
general expectation values like $t_{i_1 \ldots i_n \Phi}$ where all the 
$i_k \neq 0$.
Moreover, when one wants to compute even a very simple trace : for example
$t_{\Phi^n}$ by using \eq{eq7}, one obtains, $[{n \over 2}]$ steps later, a
$n - [{n \over 2}]$ degree  complicated trace which does not contain 
quadratic terms any more. Thus, the recursion stops there.
In fact, this problem can be overcome  by a very simple idea :
one uses the invariance of traces by cyclic permutations to get rid
of the $n+1$ degree term in \eq{eq7}. Then, one obtains relations between
general traces, and it is thus possible to compute the expectation values
of any trace in function of the first ones.
%These equations relate any expectation value of trace containing $\Phi^2$ to
%expectation values of traces of lower orders. 
%A trace which does not contain two identical matrices side by side, however,
%cannot be obtained that way.
%Moreover, beginning with $t_{\Phi^n}$, one obtains, $[{n \over 2}]$ steps
% later, a $n - [{n \over 2}]$ order trace which does not contain $\Phi^2$
%any more.
%Thus, the recurrence stops there. 
%The idea to overcome this problem is in fact very simple : it is to use
%the invariance of traces by circular permutations to get rid of the 
%$n+1$ order term in the above equation.
%Then, one obtains relations between traces without $\Phi^2$.\\

Let us see now how this idea applies to the computation of the resolvent.
We denote :
\begin{equation}
\nonumber
\omega_{i_1 i_2 \ldots i_n}={1 \over 3 N} \langle \makebox{tr} \delta_{i_1}
\Phi \delta_{i_2} \Phi \ldots \delta_{i_n} {1 \over z-\Phi} \rangle
\end{equation}
$\omega_0 =\omega$ is the usual resolvent.
Using the change of variables :
\begin{equation}
\nonumber
\Phi \rightarrow \Phi + \epsilon {1 \over z-\Phi}
\end{equation}
we obtain the equation :
\begin{equation}
\label{e10}
z\, (U+g z)\, \omega - U-g z-g \, t_{\Phi} -\omega^2 +2\, c \,\omega_{1-1}=0 
\end{equation}
Similarly,
\begin{equation}
\nonumber
 \Phi \rightarrow \Phi + \epsilon \delta_{1} \Phi \delta_{-1} {1 \over z-\Phi} 
\end{equation}
yields :
\begin{equation}
\label{e12}
z \,(U+g z)\, \omega_{1-1}-(U+g z) \, t_{\Phi} -g\, t_{1-1\Phi} -\omega \, \omega_{1-1} +c \,\omega_{1\ 0-1} + c\, \omega_{-1-1-1}=0
\end{equation}
and, by the means of similar changes in variables, we have the equations :
\begin{equation}
z\, (U+g z)\, \omega_{-1-1-1}-(U+g z)\, t_{1-1\Phi}-g\, t_{-1-1-1\Phi} +c \,\omega_{-1\ 0-1-1}+c \,\omega_{1\ 1-1-1}- \omega \, \omega_{-1-1-1} =0
\end{equation}
\begin{equation}
z\, (U+g z)\, \omega_{1\ 1-1-1}-(U+g z)\, t_{-1-1-1\Phi} -g\, t_{1\ 1-1-1\Phi} + c\, \omega_{1\ 0\ 1-1-1}+c \, \omega_{-1-1\ 1-1-1}-\omega \, \omega_{1\ 1-1-1} =0
\end{equation}

These equations alone are not sufficient to compute $\omega(z)$.
Indeed, if we intend to calculate $\omega(z)$, we generate the function 
 $\omega_{1-1}(z)$ (\eq{e10}). Then, in turn, we generate the function
 $\omega_{-1-1-1}(z)$ (\eq{e12}) and so on.

As for $\omega$ functions containing a $0$ (i.e. a $\Phi^2$ term) such as
 $\omega_{1\ 0-1}$, they are easy to deal with : we know
how to compute traces containing $\Phi^2$.
$\omega_{1\,0-1}={1 \over 3 N}\makebox{tr} \delta_1 \Phi^2 \delta_{-1} {1 \over
z -\Phi}$, will be seen as ${1 \over 3 N} \makebox{tr} \Phi^2 \delta_{-1}
{1 \over z -\Phi} \delta_1$.
Then the change in variables :
\begin{equation}
\Phi \rightarrow \Phi + \epsilon \delta_{-1} {1 \over z-\Phi} \delta_{1}
\end{equation}
yields ($\omega_{1-1} =\omega_{-1\ 1}$ for symmetry reasons)
\begin{equation}
g \, \omega_{1
 0-1}+(U+c)\, \omega_{1-1} +c \, z \, \omega-c=0
\end{equation}
and similar changes in variables lead to the equations :
\begin{equation}
g \, \omega_{-1\ 0-1-1}+U \, \omega_{-1-1-1}+c \, \omega_{1\ 0-1}+c \, z \, \omega_{1-1}
-c \, t_{\Phi}=0
\end{equation}
\begin{equation}
g \, \omega_{1\ 0\ 1-1-1}+U \, \omega_{1\ 1-1-1} +c \, z \, \omega_{-1-1-1}-c \, t_{1-1\Phi}
+c \, \omega_{-1\ 0-1-1} - t_{\Phi} \, \omega=0
\end{equation}

But, to compute  $\omega_{-1-1\ 1-1-1}$, as  mentionned in the comments to
 \eq{eq7},
we have to substract two different changes in variables and use
cyclicity of traces :
\begin{equation}
\Phi \rightarrow \Phi + \epsilon [\Phi \delta_{-1}
 \Phi \delta_{-1} (z-\Phi)^{-1}
\delta_{-1} \Phi - \delta_{-1} \Phi \delta_{-1} (z-\Phi)^{-1}
 \delta_{-1} \Phi^2] 
\end{equation}
yields :
\begin{equation}
c \, (\omega_{-1-1\ 1-1-1}+\omega_{-1\ 1-1-1-1}-\omega_{-1-1\ 0-1}-\omega_{-1\ 0\ 1\ 1-1})-\omega_{-1-1-1}+t_{\Phi} \, \omega_{1-1}=0
\end{equation}
This equation, as we know how to compute $\omega_{1-1}$, $\omega_{-1-1-1}$,
 $\omega_{-1-1\ 0-1}$ and $\omega_{-1\ 0\ 1\ 1-1}$, relates $\omega_{-1-1\ 1-1-1}$ to
 $\omega_{-1\ 1-1-1-1}$.
$$\Phi \rightarrow \Phi + \epsilon (\Phi \delta_{-1} \Phi \delta_{-1} \Phi \delta_{-1} (z-\Phi)^{-1} \Phi - \delta_{-1} \Phi \delta_{-1} \Phi \delta_{-1} (z-\Phi)^{-1} \Phi^2)$$
allows us to relate similarly 
$\omega_{-1\ 1-1-1-1}$ to $\omega_{1-1-1-1-1}$.
Then 
\begin{equation}
\Phi \rightarrow \Phi + \epsilon (\Phi \delta_{-1} 
(z-\Phi)^{-1} \delta_{1} \Phi \delta_{-1} \Phi \delta_{1}
 - \delta_{-1} (z-\Phi)^{-1} \delta_{1} \Phi \delta_{-1} \Phi \delta_{1} \Phi )
\end{equation}
allows us to relate $ \omega_{1-1-1-1-1}$ to $ \omega_{1-1\,1\,1\,1}$, and we have 
$\omega_{1-1\ 1\ 1\ 1}=\omega_{-1\ 1-1-1-1}$ as the roles of $\delta_{1}$ and $\delta_{-1}$ are
completely symmetric.\\
Finally, as a result of these operations, we have :
\begin{equation}
\omega_{-1\ 1-1-1-1}+\omega_{-1\ 1-1-1-1}=K(z)
\end{equation}
where $K(z)$ only contains easy to compute $\omega$ functions.
We can then write an equation for $\omega_{-1\ 1-1-1-1}$ and thus for 
$\omega_{-1-1\ 1-1-1}$
which only involves $\omega$ functions that we either already know or are able
to compute similarly as was done during the two first steps of the procedure.\\

That way, our set of equations is closed, and we obtain a degree five algebraic
equation for $\omega(z)$. This expression applies to general expressions
of $U$ and $c$. For the exact expression of this equation see
Appendix A.
The equation only contains four unknown parameters :
\begin{equation}
\label{parameters}
t_{\Phi}={1 \over N} \langle \makebox{tr} \Phi_{1} \rangle, \, t_{1-1\Phi}=
{1 \over N} \langle \makebox{tr} \Phi_1 \Phi_2 \rangle, \, t_{111\Phi}=
{1 \over N} \langle \makebox{tr} \Phi_1 \Phi_2 \Phi_3 \rangle, \, t_{1\ 1-1-1\Phi}
={1 \over N} \langle \makebox{tr}  \Phi_1 \Phi_2 \Phi_1 \Phi_3 \rangle
\end{equation}

These parameters are also those that would be involved if we used the
renormalisation group method \cite{BrezinZinn,higuchi,GBFD}
 to compute the Potts-3 model.
The renormalization group flows would relate the conventional
 Potts-3  to the Potts-3 + branching
interactions model, with arbitrary  $U$ and $c$; but the presence of
$t_{111\Phi}$ shows us it would also be related to the dilute Potts-3 model,
where one has a ${1 \over N} \makebox{tr} (\Phi_1+\Phi_2+\Phi_3)^3 $ term.
Finally, the $t_{1\,1-1-1\Phi}$ term shows us it may also be related to 
more complicated quartic models. \\\

We are now going to derive from our equation the critical
 behaviour and critical line
of the model when $U=1+h \makebox{tr} \Phi^2 /6$ and $c$ is a
constant. This is the most common type of extension 
of a matrix model to branching interactions.
The values of the unknown parameters given in \eq{parameters} are fixed
 by the physical constraint that the resolvent has only one physical cut
 which corresponds to the support of the eigenvalues of $\Phi$. Then, one
 can study the critical behaviour of the model.

It is easy to look for the Potts critical line.
Indeed, the scaling behaviour of the resolvent
is then, if we denote the physical cut
of $\omega$ as $[a,b]$ :
\begin{equation}
\qquad \omega(z) \sim (z-a)^{1 \over 2} \makebox{ when } z \sim a 
\makebox{ and } \omega(z) \sim (z-b)^{6 \over 5} \makebox{ when } z \sim b
\end{equation} 
The corresponding exponent $\gamma_s$ is $-{1 \over 5}$, which corresponds
to the $C={4 \over 5}$ central charge of the model.

Rather than looking for the resolvent for any values
of the coupling constants, it is easier to search for the resolvent only
on this critical line where the presence of the $6 \over 5$
exponent leads to simple conditions on the derivatives of the algebraic 
equation.

We obtain  :

\begin{equation}
\begin{array}{l}
 105 \, c^3 + 4 \,g^2 =0\\ 
2480625 \, c^2 \, (-1-4 c + 43 c^2)\, +\, 296100\, c\, (15+113 c)\, h\, -\, 692968\, h^2 =0\\
\end{array} 
\end{equation}

Let us note here that,
when $h=0$ (no branching interactions)
we recover the Potts-3 bicritical point which agrees with Daul's result 
\cite{daul}  :

$$c={2 - \sqrt{47} \over 43}\ ,\ g={\sqrt{105} \over 2} 
\left({-3+\sqrt{47} \over 41 - \sqrt{47}}\right)^{3 \over 2}$$

Thus, we have shown that the resolvent for the model of Potts-$3$ plus
 branching interactions  obeys a degree five algebraic equation. 
We have  found the critical line and  exponent of this extended model.
This extends the results of Daul \cite{daul} who had only derived
the position of the critical point and exponent of the conventional
model. 
 Finally, let us recall that, in a recent paper \cite{PZJ},
P. Zinn-Justin obtains independently algebraic equations for similar
  problems. His method, though, does not involve
 loop equations, and  is rather in the spirit of \cite{daul}. Moreover, it
does not address the problem of branching interactions and thus overlaps our
results only in the case of the conventional Potts-3 model.

\section{The Potts-$\infty$ model :}

We are now going to briefly derive the solution for the Potts-$\infty$
model, from the equations of motion point of view.
The purpose of this part is mainly to show the efficiency 
 of our method on this $c=\infty$ model. This model was
previously studied by Wexler in \cite{wexler}.

\noindent Let us denote $\Phi=
\left(
\begin{array}{ccc}
\Phi_1 &\, & 0\\
\, & \ddots &\, \\
0 &\, & \Phi_q\\
\end{array}
\right)$,\ \  and \ \ $X={(\Phi_1 + \ldots +\Phi_q) \over N} \bigotimes
{\bf 1}_{q \times q}$.\\
We shall define the Potts-q partition function as
\begin{equation}
Z=\int d\Phi e^{-N \, V(\Phi)}
\qquad
\makebox{where}
\qquad
V(\Phi)=g {\makebox{tr} \Phi^3 \over 3 N}+U {\makebox{tr} \Phi^2 \over 2 N}+c {\makebox{tr} X^2 \over 2 N}
\end{equation}
$V(\Phi)$ is of order $q$ when $q\rightarrow \infty$.
First, let us use the equations of motion to relate 
\begin{equation}
a(x)={1 \over q N} \langle \makebox{tr} {1 \over x-\Phi} \rangle \qquad 
\makebox{to} \qquad b(y)= {1 \over q N} \langle \makebox{tr} {1 \over y - X}
\rangle
\end{equation}
Let us also denote :
\begin{equation}
d(x,y)={1 \over q N} \langle \makebox{tr} {1 \over x-\Phi } {1 \over y-X} \rangle
\end{equation}

$$\Phi \rightarrow \Phi + \epsilon {1 \over x-\Phi} {1 \over y -X} \qquad
\makebox{yields }$$
\begin{equation}
(x \, (g x + U)+ c \, y  -  a(x) -  {b(y) \over q}) \ d(x,y)\, +\, g  - 
 g  y \, b(y)  - c \, a(x)  -  b(y) \, (g  x  +  U ) \, = \, 0
\end{equation}
We can get rid of $d(x,y)$ since,
when $x\, (g x+U)+c \,y-a(x)-{b(y) \over q}=0$, $d(x,y)$ remains finite, thus
$g-g y\, b(y) -c\, a(x)-b(y)\, (g x+U)=0$.
This is sufficient to relate $a(x)$ to $b(y)$.
Moreover, the value of $b(y)$ is easy to compute when $q=\infty$.\\

Let us briefly summarize this computation :
we calculate the value of $\langle \makebox{tr} X^n \rangle$
in the $q \rightarrow \infty $ limit.
 
First :
\begin{equation}
\langle \makebox{tr} X^n \rangle = \langle \makebox{tr} \Phi_1 \ldots \Phi_n \rangle + O({1 \over q})
\end{equation}
(recall that all the $\Phi_i$ play the same role).

If we now separate the first $n$ matrices from the remaining $q-n$ (with 
$q \gg n$), and suppose there is a saddle point for the eigenvalues of
${\Phi_{n+1} + \ldots +\Phi_q \over q}$, then this saddle point is
(in the $q \rightarrow \infty$ limit) independent from the matrices 
$\Phi_1\, , \, \ldots \, , \, \Phi_n$.
Then, in this limit, up to a change in variables :
$\tilde{\Phi}_k= U \Phi_k U^{-1}$, we have $n$ independent matrice
 $\tilde{\Phi}_1 \ldots \tilde{\Phi}_n$. Each of them 
has the partition function
\begin{equation}
\label{vevc}
Z_{\Lambda_{C}}=\int \  d\tilde{\Phi}_k \ e^{-N ({g \over 3}\makebox{tr} 
\tilde{\Phi}_k^3 +{U \over 2}\makebox{tr} \tilde{\Phi}_k^2 +c \makebox{tr}
\tilde{\Phi}_k \Lambda_{C})}
\end{equation}   
As $ \makebox{tr} \Phi_1 \ldots \Phi_n = \makebox{tr} \tilde{\Phi}_1 \ldots
\tilde{\Phi}_n$, we have
$\langle \makebox{tr} X^n \rangle =
\makebox{tr} \langle \tilde{\Phi}_1 \rangle_{\Lambda_{C}} \ldots 
\langle \tilde{\Phi}_n \rangle_{\Lambda_{C}} $
where $\langle \ldots \rangle_{\Lambda_{C}}$ is the expectation value
 obtained with the partition function $Z_{\Lambda_{C}}$ (cf \eq{vevc} ).\\
The matrices $\tilde{\Phi}_k\, , \ k=1,\ldots n$ all play the same part, and
 $\langle \makebox{tr} X^n \rangle = \makebox{tr} \Lambda_{C}^n$,
thus

\begin{equation}
\makebox{tr} \Lambda_{C}^n = \makebox{tr} \langle \tilde{\Phi}_1 \rangle^n 
\end{equation}

This must give us $\Lambda_{C}$, provided we calculate 
$\langle \tilde{\Phi}_1 \rangle_{\Lambda_{C}}$ in function of $\Lambda_{C}$.
This is a solvable problem, but it is much faster to note that 
\begin{equation}
\Lambda_{C} = t_{\Phi} \, {\bf 1}_{q N \times q N}
\end{equation}
is solution. Thus, ${1 \over q N} \langle \makebox{tr}
 X^n \rangle = (t_{\Phi})^n$, and $b(y)$ is simply $(y-t_{\Phi})^{-1}$.\\

This gives us immediately the solution for $a(x)$ : it obeys a second
order equation and reads :
\begin{equation}
\label{pottsinf}
a(x)={1 \over 2} (x (U + gx)+c \, t_{\Phi} -\sqrt{(x (U+g x)+c t_{\Phi})^2
-4 (U+g x+g t_{\Phi})})
\end{equation}

The Potts-$\infty$ plus branched polymers model is thus very similar
to an ordinary pure gravity model. As previously,
we compute the parameter $t_{\Phi}$ by imposing that the resolvent
$a(x)$ has only one physical cut.
The model is critical  when $a(x)$ behaves as $(x-x_0)^{3 \over 2}$,
$x_0$ being a constant,
and the critical point verifies (as in \cite{wexler}) :

\begin{equation}
g_c={1 \over 4 \sqrt{2}} \qquad \makebox{and} \qquad c_c=-{1 \over 2}
\end{equation} 

Let us note finally that
the loop equations method used here is appropriate for the renormalization
 group method of \cite{BrezinZinn,higuchi,GBFD}. 

\section{Conclusion}

In this letter, we have shown that it is possible to solve the Potts-3
and Potts-$\infty$ models on two-dimensional random lattices through
 the method of the equations of motion.
We have obtained a closed set of loop equations for the Potts-3 model,
which was thought to be impossible. We have shown that the Potts-3
resolvent obeys an order five equation, and this new knowledge opens
the door to the calculation of expectation values of the operators
of the model. We have extended the Potts-3 conventional model to Potts-3 plus
branching interactions, and given the general algebraic equation and
the Potts critical line of this model.
Finally, we have shown our method also applies successfully to
another Potts model : the Potts-$\infty$ model.
\iffalse
While we were working on these models
 with added branching interactions, though, 
we had a discussion with P. Zinn-Justin where it appeared he had another method
to solve Potts-3 (and even Potts-4) matrix model \cite{PZJ}
 with dilution terms.
He also obtains algebraic equations for the resolvents.
However, our method is different and, since  
it relies on the use of the equations of motion, it seems more adapted
to the renormalization group method \cite{higuchi,GBFD}.
Indeed, we have treated Potts + branched polymers models in order to
compare low $q$ to large $q$ critical lines and, if
possible, renormalization group flows. This
corresponds to $c < 1$  and $c >1$ (Potts-$\infty$ is a $c=+ \infty$ model)
models, and we would like to verify if F. David's conjecture \cite{FD},
applied to such models, is right.
This work is still in progress,
 in particular for large $q$
Potts + branched polymers models.
\fi
We hope to generalize soon our method to more general Potts-$q$ models,
in particular for large-$q$ Potts + branching
interactions models.
\appendix 
\section{The equation for the Potts-3 resolvent :}

Here is the degree five equation for the resolvent of this model,
where ${\bf W}(x)$ is related to $\omega(x)$ by ${\bf W}(x)=
\omega(x)- g x^2-U x$.\\\

\noindent $-24\,{c^7} + 4\,{c^4}\,{g^2} - 16\,t_{+-\phi} \,{c^5}\,{g^2}
 - 12\,t_{111\phi}
 \,{c^4}\,{g^3} - 
  4\,t_{1-1\phi} \,{c^2}\,{g^4} + 8\,t_{11-1-1\phi} \,{c^3}\,{g^4} + 
  68\,{c^6}\,g\, t_{\phi} + 2\,{c^3}\,{g^3}\, t_{\phi} + 
  3\,{c^2}\,{g^4}\, t_{\phi}^2 + 60\,{c^6}\,U + 2\,{c^3}\,{g^2}\,U - 
  52\,t_{1-1\phi} \,{c^4}\,{g^2}\,U + 20\,t_{111\phi} \,{c^3}\,{g^3}\,U - 
  20\,{c^5}\,g\, t_{\phi}\,U - 4\,{c^2}\,{g^3}\, t_{\phi}\,U - 
  36\,{c^5}\,{U^2} - 3\,{c^2}\,{g^2}\,{U^2} + 36\, t_{1-1\phi}
 \,{c^3}\,{g^2}\,{U^2} - 
  36\,{c^4}\,g\, t_{\phi} \,{U^2} - 12\,{c^4}\,{U^3} + 
  36\,{c^3}\,g\, t_{\phi}\,{U^3} + 12\,{c^3}\,{U^4} + 28\,{c^6}\,g\,x + 
  2\,{c^3}\,{g^3}\,x - 12\,t_{1-1\phi} \,{c^4}\,{g^3}\,x + 8\,t_{111\phi} 
\,{c^3}\,{g^4}\,x - 
  36\,{c^5}\,{g^2}\, t_{\phi}\,x - 2\,{c^2}\,{g^4}\, t_{\phi}\,x - 
  52\,{c^5}\,g\,U\,x - 4\,{c^2}\,{g^3}\,U\,x + 36\, t_{1-1\phi} 
\,{c^3}\,{g^3}\,U\,x - 
  18\,{c^4}\,{g^2}\, t_{\phi}\,U\,x + 2\,{c^4}\,g\,{U^2}\,x + 
  54\,{c^3}\,{g^2}\, t_{\phi }\,{U^2}\,x + 22\,{c^3}\,g\,{U^3}\,x - 
  24\,{c^5}\,{g^2}\,{x^2} - {c^2}\,{g^4}\,{x^2} + 8\, t_{1-1\phi} 
\,{c^3}\,{g^4}\,{x^2} + 
  2\,{c^4}\,{g^3}\, t_{\phi }\,{x^2}
 + 24\,{c^7}\,U\,{x^2} + 20\,{c^4}\,{g^2}\,U\,{x^2} + 
  26\,{c^3}\,{g^3}\, t_{\phi }\,U\,{x^2} - 48\,{c^6}\,{U^2}\,{x^2} + 
  12\,{c^3}\,{g^2}\,{U^2}\,{x^2} + 24\,{c^5}\,{U^3}\,{x^2} 
+ 24\,{c^7}\,g\,{x^3} + 
  6\,{c^4}\,{g^3}\,{x^3} + 4\,{c^3}\,{g^4}\, t_{\phi }\,{x^3}
 - 64\,{c^6}\,g\,U\,{x^3} + 
  2\,{c^3}\,{g^3}\,U\,{x^3}
 + 44\,{c^5}\,g\,{U^2}\,{x^3} - 16\,{c^6}\,{g^2}\,{x^4} + 
  24\,{c^5}\,{g^2}\,U\,{x^4} + 4\,{c^5}\,{g^3}\,{x^5} + 
  \left( \right. 24\,{c^5}\,g - 12\,t_{1-1\phi} \,{c^3}\,{g^3} 
+ 8\,t_{111\phi} \,{c^2}\,{g^4} - 
     36\,{c^4}\,{g^2}\, t_{\phi } - 2\,c\,{g^4}\, t_{\phi }
 - 40\,{c^4}\,g\,U - 
     2\,c\,{g^3}\,U + 36\,t_{1-1\phi} \,{c^2}\,{g^3}\,U
 - 18\,{c^3}\,{g^2}\, t_{\phi}\,U - 
     10\,{c^3}\,g\,{U^2} + 54\,{c^2}\,{g^2}\, t_{\phi }\,{U^2}
 + 26\,{c^2}\,g\,{U^3} + 
     24\,{c^7}\,x - 24\,{c^4}\,{g^2}\,x - 2\,c\,{g^4}\,x + 16\,t_{1-1\phi} 
\,{c^2}\,{g^4}\,x + 
     4\,{c^3}\,{g^3}\, t_{\phi }\,x - 36\,{c^6}\,U\,x +
 4\,{c^3}\,{g^2}\,U\,x + 
     52\,{c^2}\,{g^3}\, t-{\phi }\,U\,x + 12\,{c^5}\,{U^2}\,x + 
     36\,{c^2}\,{g^2}\,{U^2}\,x - 12\,{c^4}\,{U^3}\,x + 12\,{c^3}\,{U^4}\,x - 
     4\,{c^6}\,g\,{x^2} + 14\,{c^3}\,{g^3}\,{x^2}
 + 12\,{c^2}\,{g^4}\, t_{\phi }\,{x^2} - 
     36\,{c^5}\,g\,U\,{x^2} + 10\,{c^2}\,{g^3}\,U\,{x^2}
 + 30\,{c^4}\,g\,{U^2}\,{x^2} + 
     22\,{c^3}\,g\,{U^3}\,{x^2} - 40\,{c^5}\,{g^2}\,{x^3}
 + 60\,{c^4}\,{g^2}\,U\,{x^3} + 
     12\,{c^3}\,{g^2}\,{U^2}\,{x^3} + 18\,{c^4}\,{g^3}\,{x^4}
 + 2\,{c^3}\,{g^3}\,U\,{x^4} \left. \right) 
    \, {\bf W}(x) + \left( \right. 12\,{c^6} - 6\,{c^3}\,{g^2} + 8\,t_{1-1\phi} \,c\,{g^4} + 
     2\,{c^2}\,{g^3}\, t_{\phi } - 12\,{c^5}\,U - 4\,{c^2}\,{g^2}\,U + 
     26\,c\,{g^3}\, t_{\phi }\,U - 12\,{c^4}\,{U^2} + 18\,c\,{g^2}\,{U^2} + 
     12\,{c^3}\,{U^3} - 44\,{c^5}\,g\,x - 2\,{c^2}\,{g^3}\,x + 
     12\,c\,{g^4}\, t_{\phi }\,x + 30\,{c^4}\,g\,U\,x + 20\,c\,{g^3}\,U\,x + 
     26\,{c^2}\,g\,{U^3}\,x + 6\,{c^4}\,{g^2}\,{x^2} + 2\,c\,{g^4}\,{x^2} + 
     12\,{c^3}\,{g^2}\,U\,{x^2} +
 39\,{c^2}\,{g^2}\,{U^2}\,{x^2} + 20\,{c^3}\,{g^3}\,{x^3} + 
     14\,{c^2}\,{g^3}\,U\,{x^3} + {c^2}\,{g^4}\,{x^4}\left. \right)
 \,{{\bf W}(x)^2} + 
  \left( \right. -12\,{c^4}\,g - 2\,c\,{g^3} + 4\,{g^4}\, t_{\phi } - 10\,{c^3}\,g\,U + 
     4\,{g^3}\,U + 26\,{c^2}\,g\,{U^2} + 20\,{c^3}\,{g^2}\,x 
+ 4\,{g^4}\,x + 
     12\,{c^2}\,{g^2}\,U\,x + 18\,c\,{g^2}\,{U^2}\,x 
+ 22\,c\,{g^3}\,U\,{x^2} + 4\,c\,{g^4}\,{x^3}
    \left.  \right) \,{{\bf W}(x)^3} + \left( \right. -( {c^2}\,{g^2} ) 
 + 18\,c\,{g^2}\,U + 
     4\,c\,{g^3}\,x + 4\,{g^3}\,U\,x + 4\,{g^4}\,{x^2}\left.  \right) 
\,{{\bf W}(x)^4} + 4\,{g^3}\,{{\bf W}(x)^5} = 0$\\

Note that $U$ and $c$ may depend, in the most general case, on
$t_{1-1\Phi}$ and  $t_{\Phi^2}$, the latter being related to $t_{\Phi}$
through the equation of motion : $g \, t_{\Phi^2} + (U+2 \, c) \, t_{\Phi}=0$.
In this article, we have computed explicitely the critical line for
 the particular case of $c$ constant and $U=1 + {h \over 2} \, t_{\Phi^2}$.

\section{Acknowledgments :}

We thank P. Zinn-Justin for discussing his ideas with us,
 and we are grateful to
F. David and J.-B. Zuber 
for useful discussions and careful reading of the manuscript.

\end{document}